\documentclass[twoside,openright,a4paper]{article}
\usepackage{graphicx}
\usepackage{graphics}
\usepackage{epsfig}
\usepackage{amsmath}
\usepackage{amsfonts}
\usepackage{sectsty}
\sectionfont{\large}
.360pk
\font\rt=cmss9.360pk
\font\sd=cmcsc9.360pk

\begin{document}

\centerline{\Large Three-Dimensional Smoothed Particle Hydrodynamics}
\centerline{\Large Method for Simulating Free Surface Flows}

\medskip

\begin{center}
{\sc Rizal Dwi Prayogo$^{a,b}$, Christian Fredy Naa$^a$}

\medskip

$^a$Faculty of Mathematics and Natural Sciences, Institut Teknologi Bandung, Jl. Ganesha 10, Bandung 40132 Indonesia, E-mail: rizal.dp@s.itb.ac.id, chris@cphys.fi.itb.ac.id
  \\ 
$^b$Graduate School of Natural Science and Technology, Kanazawa University, Kakuma, Kanazawa 920-1192 Japan,\\
\end{center}

\bigskip

{\parindent 0pt
\textbf{Abstract.} {\em In this paper, we applied an improved Smoothing Particle Hydrodynamics (SPH) method by using gradient kernel renormalization in three-dimensional cases. The purpose of gradient kernel renormalization is to improve the accuracy of numerical simulation by improving gradient kernel approximation. This method is implemented for simulating free surface flows, in particular dam break case with rigid ball structures and the propagation of waves towards a slope in a rectangular tank.}\\
\newline
\textbf{Keywords:} Smoothed particle hydrodynamics, free surface flows, gradient kernel renormalization
}


\section{Introduction}
Computational Fluid Dynamics (CFD) using Smoothed Particle Hydrodynamics (SPH) has a wide range of applications to solve problem in engineering and science. SPH is a mesh-free Lagrangian method and well suited to the simulation of complex and free surface flows. The SPH method was originally used to model astrophysical problems by Lucy \cite{Lucy1977} and Gingold and Monaghan \cite{Gingold1977}. Three-dimensional SPH method has been studied numerically by Monaghan \cite{Monaghan1992} in a field of astrophysical fluid dynamics processes.

We obtain the SPH equations from the continuum equations of fluid dynamics by interpolating from set of points which may be disordered. This interpolation is based on the theory of integral interpolants using interpolation kernels which approximate the delta function. The interpolants being analytic functions can be differentiated without using grids. 

Monaghan \cite{Monaghan1994} studied the application of the particle method SPH to free surface problems in two-dimensional cases. In this paper, we applied an improved SPH method by using gradient kernel renormalization in three-dimensional cases. The purpose of gradient kernel renormalization is to improve the accuracy of the simulations \cite{Oger2007}.

In the following, first the general concept of SPH method is given. The improved SPH method using gradient kernel renormalization is introduced and described in detail. This improved SPH method is implemented for simulating free surface flows, in particular dam break case with rigid ball structures and the propagation of waves towards a slope in a rectangular tank.


\section{Smoothed particle hydrodynamics}
The SPH equations are described in detail by Liu and Liu \cite{Liu2003}. In this paper, we consider the application of three-dimensional SPH to free surface problems. The SPH method represents continuous fluid using a set of particles. Each particle $i$ has physical quantities, such as mass $m_i$, position $\mathbf{r}_i$, velocity $\mathbf{v}_i$, density $\rho_i$, and pressure $P_i$. Each particle in the SPH method is associated with a support domain. The SPH approximation, which consists of the particle approximation and the kernel approximation, is performed within the current support domain. The value of a function defining a physical quantity can be approximated by its values at a number of neighboring particles. The SPH method uses the concept of integral representation of a field function $f(x)$ by the following identity
\begin{equation}
<f(\mathbf{r})> = \int_\Omega f(\mathbf{r'})W(|\mathbf{r}-\mathbf{r'}|,h)d\mathbf{r'},
\label{eq:intrep}
\end{equation}
where $\mathbf{r}$ and $\mathbf{r}'$ are the position vectors, $W$ is the smoothing function or kernel function, and $h$ is the smoothing length defining the influence radius of $W$.
\begin{figure}[h]
\centering
\includegraphics[scale=0.4]{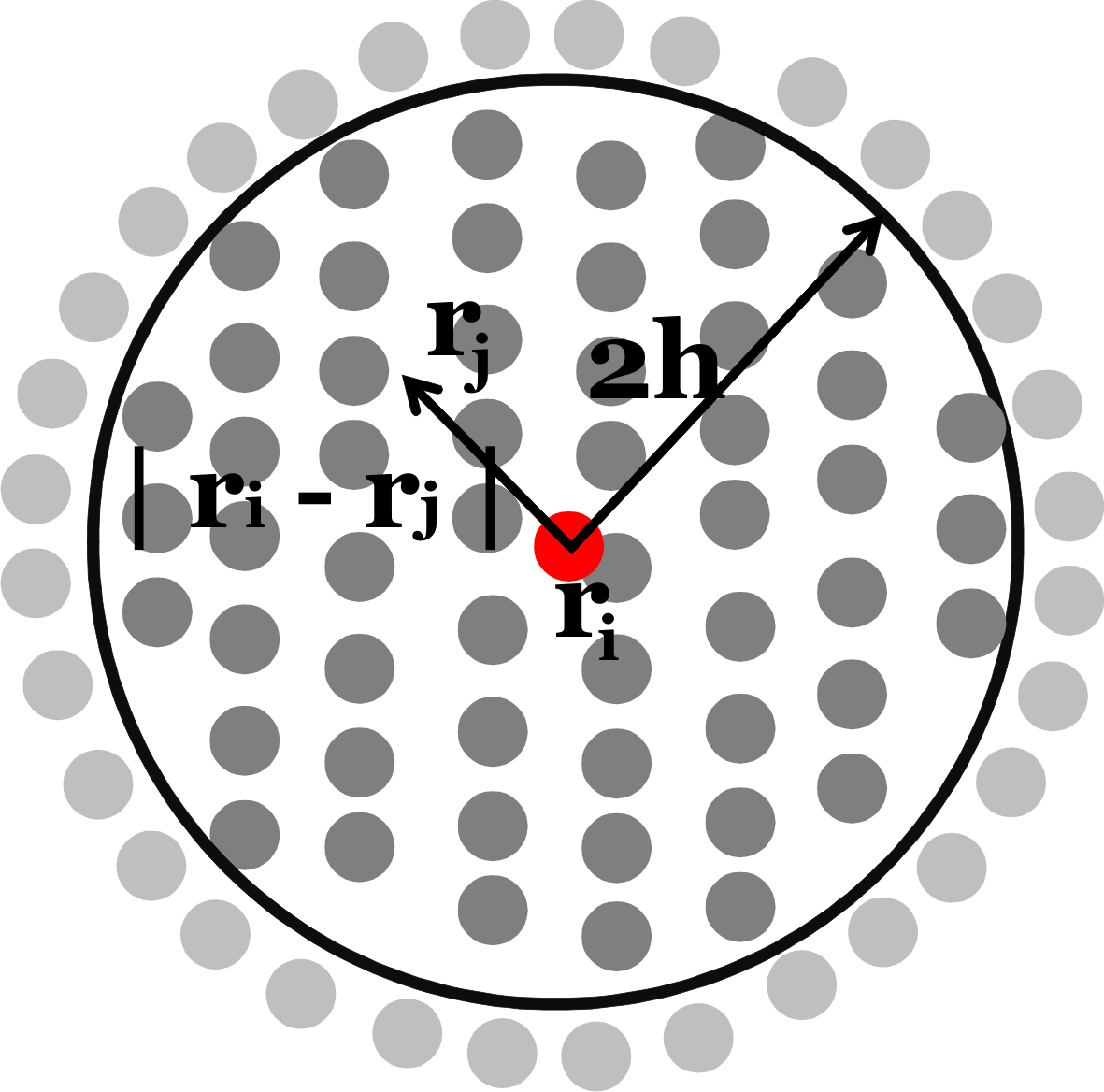}
\caption{The influence radius of $W$.}
\end{figure}

In SPH approximation, there are various kernel functions. Since it is affects the accuracy and stability of numerical results, the choice of kernel function $W$ is important to consider. The integral representation should satisfies several conditons. That is the normalization condition
\begin{equation*}
\int_\Omega W(|\mathbf{r}-\mathbf{r'}|,h)d\Omega = 1,
\end{equation*}
delta function property
\begin{equation*}
\lim_{h\to 0} W(|\mathbf{r}-\mathbf{r'}|,h)=\delta (|\mathbf{r}-\mathbf{r'}|),
\end{equation*}
moreover, often the compact support condition is required
\begin{equation*}
W(|\mathbf{r}-\mathbf{r'}|,h) = 0 \text{ outside of support domain.}
\end{equation*}

We converted the continuous integral representation (\ref{eq:intrep}) into discretized forms as a summation over all the particles in the support domain. This process is also commonly known as particle approximation in the SPH literature \cite{Liu2003}. Writing the particle approximation as follows
\begin{equation*}
<f(\mathbf{r})>=\sum_{j=1}^{N}\dfrac{m_j}{\rho_j}f(\mathbf{r_j})W(|\mathbf{r}-\mathbf{r_j}|,h),
\end{equation*}
where $m_j$ and $\rho_j$ are the mass and density of the particle $j$, respectively, and $j=1, 2,\ldots, N$, where $N$ is the total number of neighboring particles in the influence domain $\Omega$. In this paper, we use cubic spline kernel as follows \cite{Monaghan1985}
\begin{equation*}
W (q,h) = \dfrac{1}{\pi h^3} 
\begin{cases}
1-\frac{3}{2} q^2+\frac{3}{4} q^3, & \text{$0\leq q < 1$}\\
\frac{1}{4}(2-q)^3, & \text{$1\leq q < 2$}\\
0, & \text{otherwise,}
\end{cases}
\end{equation*}
where $q=\dfrac{|\mathbf{r_i}-\mathbf{r_j}|}{h}$ is the relative distance of particle $i$ and $j$.


\section{Numerical model}
\subsection{The continuity equation}
The continuity equation is based on the conservation of mass. We write the continuity equation in the form
\begin{equation}
\dfrac{D\rho}{Dt}=-\rho\nabla\mathbf{v},
\label{eq:continuity}
\end{equation}
where $\mathbf{v}$ and $\rho$ are velocity and density, respectively. Writing (\ref{eq:continuity}) in SPH discretization form as in \cite{Monaghan1988}, we obtain
\begin{equation}
\dfrac{D\rho_i}{Dt}=\rho_i\sum_j \dfrac{m_j}{\rho_j}(\mathbf{v_i}-\mathbf{v_j})\nabla_{r_i}W(|\mathbf{r_i} - \mathbf{r_j}|,h), 
\label{eq:continuitydiscrete}
\end{equation}
where $\rho_k$ and $\mathbf{v}_k$ are density and velocity of particle $k$ (evaluated at $k=i$ or $k=j$), respectively, $m_j$ is mass of particle $j$ and
\begin{equation*}
	\nabla_{r_i}W(|\mathbf{r_i}-\mathbf{r_j}|,h)=\dfrac{\mathbf{r_i}-\mathbf{r_j}}{|\mathbf{r_i}-\mathbf{r_j}|}\dfrac{\partial W}{\partial r}.
\end{equation*}

\subsection{The momentum equation}
The momentum equation is based on the conservation of momentum which is given by
\begin{equation}
\dfrac{D\mathbf{v}}{Dt} = -\dfrac{1}{\rho}\nabla P + \mathbf{F},
\label{eq:momentum}
\end{equation}
where $\mathbf{v}, \rho$ and $P$ are velocity, density, and pressure, respectively. Here, $F$ is external force, in this case gravitational acceleration. Writing (\ref{eq:momentum}) in SPH discretization form as in \cite{Monaghan1988}, we get
\begin{equation}
\dfrac{D\mathbf{v_i}}{Dt} = -\sum_j m_j\Big(\dfrac{P_i+P_j}{\rho_i \rho_j}+\Pi_{ij}\Big)\nabla_{r_i} W(|\mathbf{r_i} - \mathbf{r_j}|,h) + \mathbf{F},
\label{eq:momentumdiscrete}
\end{equation}
where $P_k$ is pressure of particle $k$ (evaluated at $k=i$ or $k=j$). 

In SPH, there are various formulations for viscosity. In the momentum equation, the introduction of a viscous term is necessary not only to consider viscid fluids and no slip boundary conditions, but also to provide the stability to the system and to prevent inter-particle penetration. The artificial viscosity term $\Pi_{ij}$ is added to pressure terms within the momentum equation (\ref{eq:momentumdiscrete}). The artificial viscosity $\Pi_{ij}$ has the form \cite{Monaghan1992}
\begin{equation*}
\Pi_{ij} = 
\begin{cases}
\dfrac{-\alpha c\mu_{ij}+\beta \mu_{ij}^2}{(\rho_i + \rho_j)/2}, & \text{$(\mathbf{v_{i}-\mathbf{v_j}}).(\mathbf{r_{i}-\mathbf{r_j}})<0$}\\
0, & \text{$(\mathbf{v_{i}-\mathbf{v_j}}).(\mathbf{r_{i}-\mathbf{r_j}})>0$}
\end{cases}
\end{equation*}
where 
\begin{align*}
\mu_{ij} = \dfrac{h(\mathbf{r_{i}-\mathbf{r_j}}).(\mathbf{v_{i}-\mathbf{v_j}})}{|\mathbf{r_i} - \mathbf{r_j}|^2 + \eta^2}.
\end{align*}

In these expressions, $c$ is the speed of sound, $\eta = 0.001$, $\alpha$ and $\beta$ represent shear and bulk viscosity, respectively.  For the problems described here, we choose $\alpha = 0.03$ and $\beta = 0$.

\subsection{The equation of state}
The equation of state is used to relate density to pressure. In this paper, the Tait's equation of state has the form
\begin{equation*}
p=\dfrac{\rho_0c^2}{\gamma}\Big[\Big(\dfrac{\rho}{\rho_0}\Big)^\gamma-1\Big],
\end{equation*}
where $c, \rho_0$, and $\gamma$ are the speed of sound, density reference, and the polytropic constant, respectively. Note that $\gamma = 7$ is usually used for water simulations. The speed of sound $c$ is approximately $\sqrt{100gH}$ and it is chosen in respect of a low Mach number ($Ma<0.1$) to ensure low compressibility effects \cite{Monaghan1992}.

\section{Improvement of the SPH method}
In this paper, we applied an improvement to the standard SPH method by using renormalization. This technique is to improve the accuracy of the method \cite{Oger2007}.

\subsection{Gradient kernel renormalization}
The velocity gradient in \eqref{eq:continuitydiscrete} can be approached by using
\begin{align*}
	\nabla\mathbf{v}=\nabla\mathbf{v}-\mathbf{v}\nabla 1.
\end{align*}

We can generalize this approach for any field $f$ by using
\begin{equation}
	\nabla f = \nabla f- f\nabla 1
	\label{eq:generalize}
\end{equation}
and transforming (\ref{eq:generalize}) into its continuous convoluted form we have
\begin{equation*}
		\langle\nabla f(\mathbf{r})\rangle = \int_\Omega f(\mathbf{r'})\nabla W d\mathbf{r'} - f(\mathbf{r})\int_\Omega \nabla W d\mathbf{r'}.
\end{equation*}

We recall the second order Taylor expansion 
	\begin{align*}
		\int_\Omega f(\mathbf{r'})\nabla W d\mathbf{r'} &=  f(\mathbf{r})\int_\Omega \nabla W d\mathbf{r'} + 
		\dfrac{\partial f(\mathbf{r})}{\partial \mathbf{r_1}}\underbrace{\int_\Omega (\mathbf{r_1'}-\mathbf{r_1})\nabla W d\mathbf{r'}}_A +\\
		&\dfrac{\partial f(\mathbf{r})}{\partial \mathbf{r_2}}\underbrace{\int_\Omega (\mathbf{r_2'}-\mathbf{r_2})\nabla W d\mathbf{r'}}_B +
		\dfrac{\partial f(\mathbf{r})}{\partial \mathbf{r_3}}\underbrace{\int_\Omega (\mathbf{r_3'}-\mathbf{r_3})\nabla W d\mathbf{r'}}_C + O(h^2).
	\end{align*}
	
In order to ensure gradient interpolations of linear fields, it is necessary to ensure that the discrete approximation of $A$, $B$, and $C$ are		
	\begin{align*}  		
  		A = 
		\begin{pmatrix}
			1 \\ 0 \\ 0
		\end{pmatrix}
		\quad
		B = 
		\begin{pmatrix}
			0 \\ 1 \\ 0
		\end{pmatrix}
		\quad
		C = 
		\begin{pmatrix}
			0 \\ 0 \\ 1
		\end{pmatrix}.
	\end{align*}
	
By the renormalization procedure \cite{Oger2007}, we modify $\nabla W$ as follows
\begin{align*}
		\sum_j \dfrac{m_j}{\rho_j}(\mathbf{r_j}-\mathbf{r}) \footnotesize{\bigotimes} L(\mathbf{r})\nabla W(|\mathbf{r}-\mathbf{r}_j|)=
		\begin{pmatrix}
			1 & 0 & 0\\
			0 & 1 & 0\\
			0 & 0 & 1
		\end{pmatrix},
	\end{align*}
	where $L(\mathbf{r})$ is a $(d,d)$ correction matrix, $d$ is the dimension of the case. In this paper, we consider three-dimensional cases ($d=3$) and calculate $L(\mathbf{x})$ to increase the accuracy of gradient kernel approximation. The continuity equation is discretized by the following manner
\begin{equation*}
\dfrac{D\rho_i}{Dt}=\rho_i\sum_j \dfrac{m_j}{\rho_j}(\mathbf{v_i}-\mathbf{v_j})L(\mathbf{r_i})\nabla_{r_i}W(|\mathbf{r_i} - \mathbf{r_j}|,h). 
\end{equation*}

This discretized form ensures exact interpolations for both constant and linear fields. Note that we can discretize the conservation of momentum by the following manner
\begin{equation*}
\dfrac{D\mathbf{v_i}}{Dt} = -\sum_j m_j\Big(\dfrac{P_i+P_j}{\rho_i \rho_j}+\Pi_{ij}\Big)L(\mathbf{r_i})\nabla_{r_i} W(|\mathbf{r_i} - \mathbf{r_j}|,h) + \mathbf{F}.
\end{equation*}


\subsection{Numerical time integration with renormalization}
As the other explicit hydrodynamic methods, different numerical time integrations can be applied in SPH simulation, such as Leap-Frog, predictor-corrector, Runge-Kutta, and Beeman schemes. The advantages of the Leap-Frog algorithm are its low memory usage on storage and its computational efficiency. We applied it in this paper with its improvement by using gradient kernel renormalization. Therefore,
	\begin{align*}
		\mathbf{r_i^*} = \mathbf{r_i^n} + \dfrac{dt}{2}\mathbf{v_i^n},
	\end{align*}
	\begin{align*}
		p_i^n=\dfrac{\rho_0c_0^2}{\gamma}\Big[\Big(\dfrac{\rho_i^n}{\rho_0}\Big)^\gamma-1\Big],
	\end{align*},
	\begin{align*}
		\mathbf{v}_i^{n+1/2} = \mathbf{v}_i^{n-1/2}-dt\sum_j m_j\Big(\dfrac{p_i^n+p_j^n}{\rho_i^n \rho_j^n}+\Pi_{ij}\Big)L(r_i^{n+1/2})\nabla_{r_i} W(\mathbf{r}_{ij}^{n+1/2},h) + \mathbf{F}_i^{n+1/2},
	\end{align*}
	\begin{align*}
		\rho_i^{n+1/2}=\rho_i^{n-1/2} + dt\rho_i \sum_j \dfrac{m_j}{\rho_j}\big(\mathbf{v_i^{n+1/2}}-\mathbf{v_j^{n+1/2}}\big)L(r_i^{n+1/2})\nabla_{r_i} W(\mathbf{r}_{ij}^{n+1/2},h),
	\end{align*}
	\begin{align*}
	\mathbf{r}_i^{n+1}=\mathbf{r}_i^{n+1/2}+\dfrac{dt}{2}\mathbf{v}_i^{n+1/2}.
	\end{align*}

\newpage
\section{Implementation}
In the following section, the results of numerical simulations for improved SPH method are given. This method was implemented in dam-break problem with rigid ball structures and water waves generated by oblique piston type wave-maker.

\subsection{Dam-break and structure}
\begin{figure}[h]
\centering
	\includegraphics[scale=0.385]{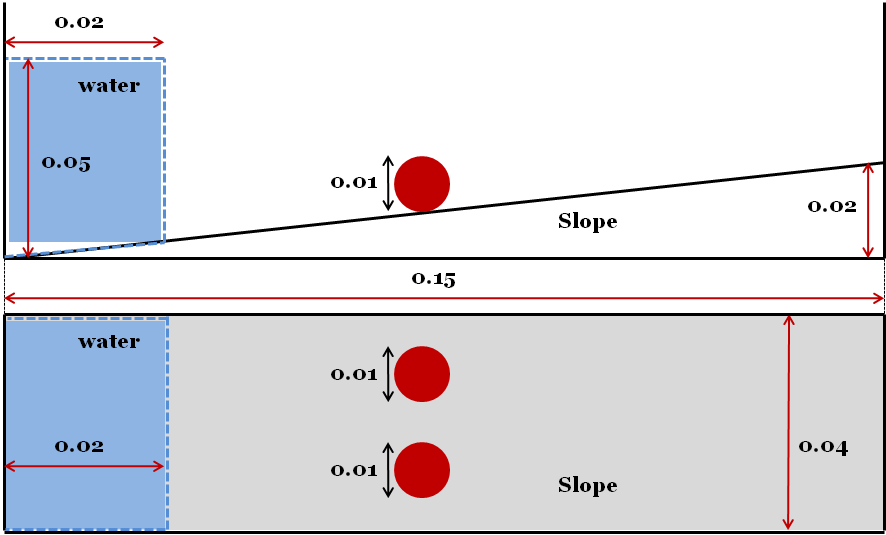}
	\caption{Sketch of dam-break problem considered, side view (upper panel) and top view (lower panel).}
	\label{fig:rigidball}
\end{figure}

In this implementation, we consider a rectangular tank with three-dimensional problem, in particular on interaction between waves and structures. Here we examine the impact of a single wave with rigid ball structures over the slope by means of a three-dimensional SPH method. A rectangular tank contains fixed structures and we used 10075 particles for this simulation. The geometry is shown in Fig. \ref{fig:rigidball}.

\begin{figure}[h]
\centering
	\includegraphics[scale=0.55]{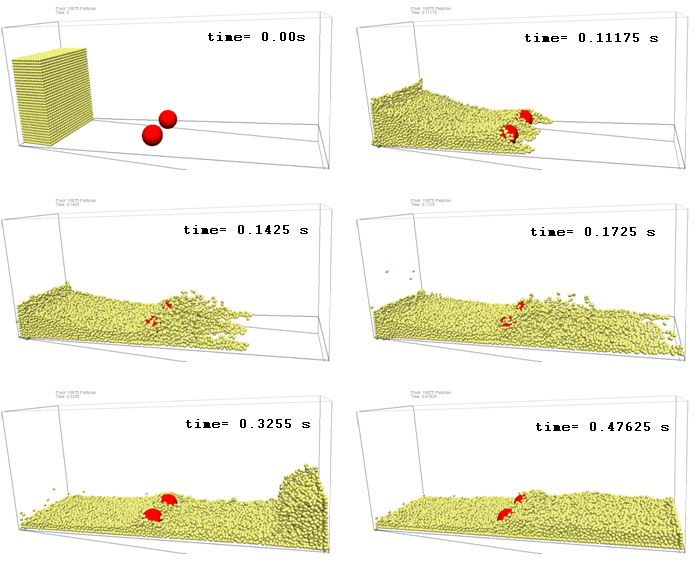}
	\caption{($T=0.0$ s) initial configuration; ($T=0.11175$ s) wave hitting the rigid ball structures; ($T=0.1425$ s) wave wrapping around the rigid ball structures; ($T=0.1725$ s) waves colliding after passing the rigid ball structures; ($T=0.3255$ s) wave colliding with the opposite wall of the tank; ($T= 0.47625$ s) reflected wave hitting the back of the rigid ball structures.}
	\label{fig:result1}
\end{figure}

Fig. \ref{fig:result1} shows the motion of a single wave which moves through rigid ball structures in a rectangular tank. The frame at $T=0.0$ s shows the initial configuration. In the next frame at $T=0.11175$ s, the wave generated by the dam break and the initial layer of water on the bottom collides with the front of the rigid ball structures and at $T=0.1425$ s the wave wraps around the rigid ball structures. At $T=0.1725$ s, the waves collide from both sides of the rigid ball structures then continue moving toward the right vertical wall. The wave reflects after colliding with the opposite wall of the tank at $T=0.3255$ s. The last movement, at $T=0.47625$ s, the reflected wave hits the back of rigid ball structures.

\newpage
\subsection{Oblique piston type wave-maker}
\begin{figure}[!h]
\centering
	\includegraphics[scale=0.385]{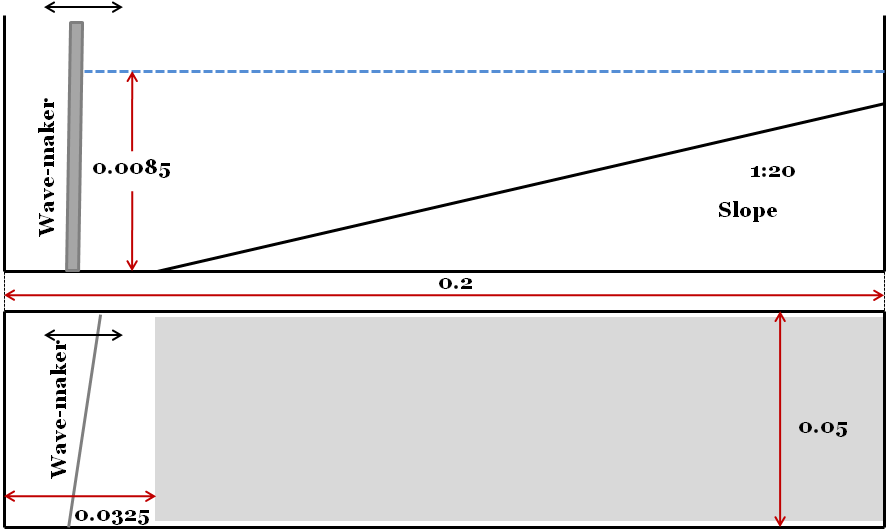}
	\caption{Sketch of oblique piston type wave-maker, side view (upper panel) and top view (lower panel).}
	\label{fig:oblique}
\end{figure}

In this implementation, we consider a rectangular tank with three-dimensional problem, in particu\-lar the propagation of waves towards a slope. This simulation involves a wave-maker in the form of an oscillating oblique piston on the left-hand side and used 80682 particles. The geometry is shown in Fig. \ref{fig:oblique}.

\begin{figure}[!h]
	\centering
	\includegraphics[scale=0.55]{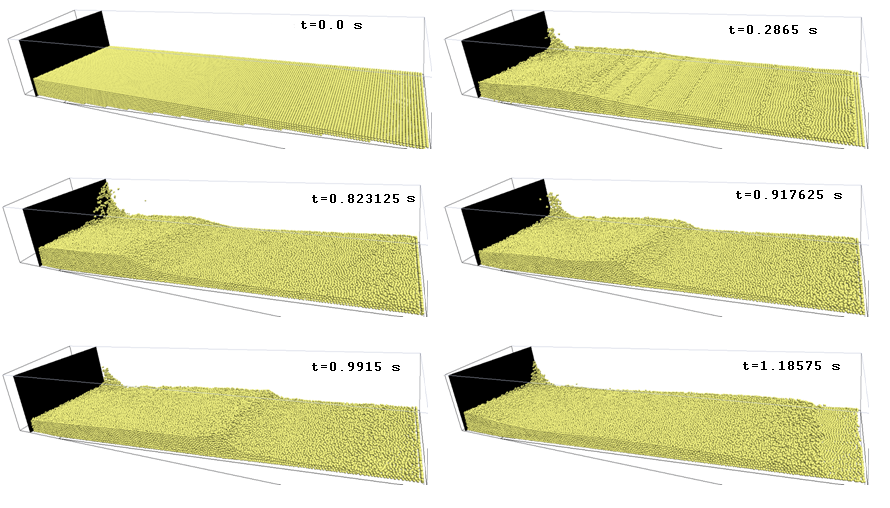}
	\caption{($T=0.0$ s) initial configuration; ($T=0.2865$ s) the wave is generated by the oscillating oblique piston type wave-maker; at ($T=0.823125$ s), ($T=0.917625$ s), and ($T=0.9915$ s) the water wave propagates onto the slope; ($T=1.18575$ s) the water wave reaches shallow water area and hits the right wall.}
	\label{fig:result2}
\end{figure}

Fig. \ref{fig:result2} shows the simulation of free surface flows. The water waves generated by oscillating piston type wave-maker were simulated. In Fig. \ref{fig:result2} the waves are shown propagating onto the slope. The frame at $T=0.0$ s shows the initial configuration with water lying on the slope. In the next frame at $T=0.2865$ s, the wave is generated by the oscillating oblique piston type wave-maker. At $T=0.823125$ s, $T=0.917625$ s, and $T=0.9915$ s the water wave propagates onto the slope. Finally, at $T=1.18575$ s the water wave reaches shallow water area and hits the right wall.


\section{Summary}
This paper presents the application of an improved SPH method by using gradient kernel renormalization for simulating free surface flows. We have implemented this method on three-dimensional cases, in particular on the interaction between waves and structures; and the propagation of waves towards a slope for waves generated by oblique piston type wave-maker. The three-dimensional case of the model has been shown to produce three-dimensional phenomenon, i.e., the collision of a single wave with rigid ball structures and its passing around the obstacle. In summary, an improved SPH model based on renormalization can be successfully used to simulate three-dimensional wave problems.

\end{document}